\documentclass[Physsubmission, Phys]{SciPost}

\hypersetup{
    colorlinks,
    linkcolor={red!50!black},
    citecolor={blue!50!black},
    urlcolor={blue!80!black}
}

\usepackage[bitstream-charter]{mathdesign}
\usepackage{slashed}
\usepackage{euscript}
\DeclareSymbolFont{usualmathcal}{OMS}{cmsy}{m}{n}
\DeclareSymbolFontAlphabet{\mathcal}{usualmathcal}
\newcommand{\df}{\mathrm{d}}

\newcommand{\X}{{\EuScript X}}
\newcommand{\G}{{\EuScript G}}

\newcommand{\Gsl}{\rlap{\hspace{0.2mm}/}{\G}}
\newcommand{\nsl}{\rlap{\hspace{0.25mm}/}{n}}

\begin{document}
\begin{center}{\Large \textbf{
			The Two-Loop Radiative Gluon Jet Function for $gg\to h$ via a Light Quark Loop\\
}}\end{center}
\begin{center}
Marvin Schnubel\textsuperscript{1$\star$}
\end{center}

\begin{center}
{\bf 1} PRISMA$^+$ Cluster of Excellence \& Mainz Institute for Theoretical Physics\\
Johannes Gutenberg University, 55099 Mainz, Germany\\
* maschnub@uni-mainz.de
\end{center}

\begin{center}
\today
\end{center}


\definecolor{palegray}{gray}{0.95}
\begin{center}
\colorbox{palegray}{
  \begin{tabular}{rr}
  \begin{minipage}{0.1\textwidth}
    \includegraphics[width=35mm]{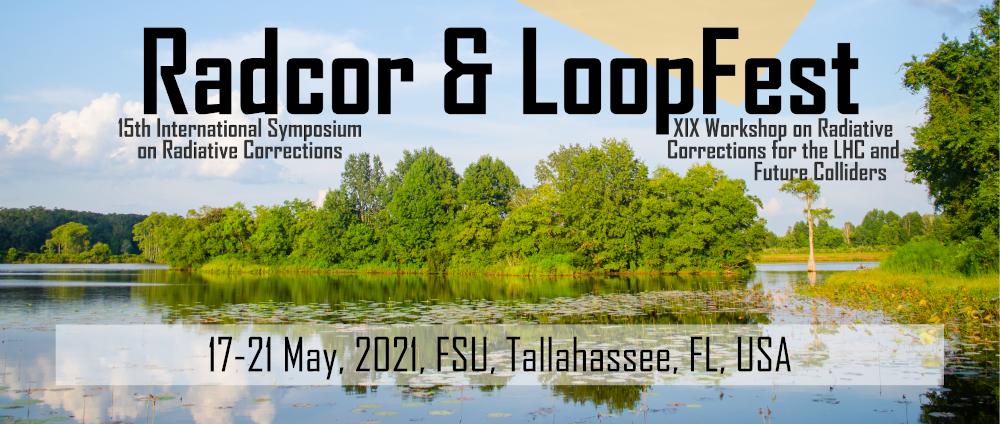}
  \end{minipage}
  &
  \begin{minipage}{0.85\textwidth}
    \begin{center}
    {\it 15th International Symposium on Radiative Corrections: \\Applications of Quantum Field Theory to Phenomenology,}\\
    {\it FSU, Tallahasse, FL, USA, 17-21 May 2021} \\
    \doi{10.21468/SciPostPhysProc.?}\\
    \end{center}
  \end{minipage}
\end{tabular}
}
\end{center}

\section*{Abstract}
{\bf
The Higgs-boson production channel $gg\to h$ mediated by light-quark loops receives large logarithmic corrections in QCD, which can be resummed using factorization formulae derived in soft-collinear effective theory. In these factorization formulae the radiative gluon jet function appears, which is a central object in the study of factorization beyond the leading order in scale ratios. We calculate this function at two-loop order for the first time and present the subtleties that come along with this.
}

\vspace{10pt}
\noindent\rule{\textwidth}{1pt}
\tableofcontents\thispagestyle{fancy}
\noindent\rule{\textwidth}{1pt}
\vspace{10pt}

\section{Introduction}
\label{sec:intro}
In the upcoming years future experiments and data analysis techniques are going to reach unprecedented accuracy in measurements of fundamental quantities. To test if our models are valid theories beyond leading approximation, it is therefore necessary to attain higher precision also in theoretical predictions. As a consequence, in the past years, the desire to understand processes such as the important Higgs production channel $gg\to h$ via two gluons at subleading order has grown. Only recently it has been shown that in the related decay process $h\to\gamma\gamma$ via a light quark loop factorization at subleading power can be achieved with the help of soft-collinear effective theory (SCET). In general, SCET has shown to be a helpful tool to study various features of problems sensitive to multiple high energy scales, because it allows the derivation of factorization theorems consisting of functions that only depend on one scale respectively \cite{Liu:2019oav,Liu:2020wbn,Bosch:2003fc,Moult:2019mog}. These functions are hard, jet and soft functions typically receiving contributions from different momentum regions in Feynman integrals. Hard functions are the Wilson coefficients obtained by matching SCET onto the full theory, whereas jet and soft functions are operator matrix elements of collinear and soft fields in the effective theory. The important jet functions are non-local products of collinear fields and often live at an intermediate scale between the hard scattering scale and the lowest one.\\
In both cases of Higgs decay $h\to\gamma\gamma$ and production $gg\to h$ via a light b-quark loop in the factorization theorem a radiative photon (gluon) jet function is encountered \cite{Liu:2020ydl}. In more detail, we match the effective SCET-II theory onto full QCD. One of the relevant operators is then further factorized into a convolution of radiative jet and soft quark soft functions. Due to the non-abelian nature of $gg\to h$ the radiative gluon jet function exhibits a more complicated color structure than the photon jet function and thus can not easily be deduced.\\
The importance of the gluon jet function is presented in \cite{HiggsGluGlu}. As is shown there, it is not only one part of the factorization theorem, but plays a vital role in refactorization theorems, which themselves are a necessity to elegantly cure divergences from endpoints of convolution integrals. Furthermore the jet function is crucial to predict large logarithmic terms in the 3-loop amplitude from RG equations.

\section{General properties of the radiative jet function}
\label{sec:properties}
We begin by setting up the definition of the jet function based on work previously done for the photon jet function due to the similarity of both quantities. Following the notation of \cite{Liu:2019oav, HiggsGluGlu}, the radiative gluon jet function is defined as
\begin{equation}
	\label{eq:jetdefinition}
	\int \df^Dy\, e^{ip_s^-\cdot y}\Big\langle g(k,a)\left|\,\mathrm{T}\left\{\,\X_{n}^{\beta k}(0) 
	\big[\bar\X_{n}(y)\,\Gsl_{n}^\perp(y)\big]^{\gamma l}\right\}\,
	\right|0\Big\rangle = g_sT^a_{kl}\,\Big[\,\frac{\nsl}{2}\,\rlap/\varepsilon_\perp^*(k) \Big]^{\beta\gamma}
	\,\frac{i\bar n\cdot p}{p^2+i0}\,J_g\big(p^2\big)\,,
\end{equation}
where
\begin{equation}
	\label{eq:kinematics}
	k=\bar{n}\cdot k\frac{n}{2}=\frac{1}{2}m_H n\,,~p_s^+=n\cdot p_s\frac{\bar{n}}{2}=\frac{1}{2}\lambda_s m_H \bar{n}\,,
\end{equation}
and $n$ and $\bar{n}$ are the light-like reference vectors. Choosing the frame as the Higgs rest frame, they are defined as $n=(1,0,0,1)^T$ and $\bar{n}=(1,0,0,-1)^T$ and fulfill $n\cdot\bar{n}=2$. The gauge-invariant building blocks $\X$ and $\G$ are defined in terms of collinear quark spinors $\xi_n$ and gluon fields $G$ as 
\begin{eqnarray}
	\label{eq:gaugeinv}
	\begin{aligned}
		\X(x)&=W^\dagger(x)\xi_n(x)\\
		\G_\perp^\mu(x)&=W^\dagger(x)\left(iD_{n\perp}^\mu W(x)\right)\,,
	\end{aligned}
\end{eqnarray}
where in the second line $D_{n\perp}^\mu$ is the $\perp$-component of a covariant collinear derivative acting on collinear fields, $W_n$ is a collinear Wilson line\footnote{For a more detailed introduction into the factorization of $gg\to h$ see \cite{HiggsGluGlu}. A pedagogical introduction to SCET is given in \cite{Becher:2014oda}} \cite{Bauer:2002nz, Hill:2002vw}.\\
For our calculation we will work in the so-called light-cone gauge $\bar{n}\cdot G=0$. This dramatically simplifies the expressions for Wilson lines and gauge-invariant building blocks, but comes at the cost of introducing a more complicated gluon propagator. In more detail, we find that all Wilson lines are equal to $1$ and $\G_n^\mu(x)=g_s G_n^\mu(x)$, and the expression for the propagator yields
\begin{equation}
	P^{\mu\nu}(l)=\frac{-i}{l^2+i0}\left[g^{\mu\nu}-\frac{\bar{n}^\mu l^\nu+\bar{n}^\nu l^\mu}{\bar{n}\cdot l}\right]\,.
\end{equation}
Additionally, no ghosts contribute to the diagrams. To check for consistency, all the calculations have been redone in a covariant gauge. To project out the Dirac structure of the jet function we make use of the following helpful identity
\begin{equation}
	\label{eq:projector}
	\mathrm{Tr}\left[\frac{\slashed{n}}{2}\gamma^\mu_\perp\gamma_\perp^\nu\frac{\slashed{\bar{n}}}{2}\right]=2g_\perp^{\mu\nu},\,\, \text{with}\,\,g_\perp^{\mu\nu}=g^{\mu\nu}-\frac{n^\mu\bar{n}^\nu+n^\nu\bar{n}^\mu}{2}.
\end{equation}
Since the collinear SCET Lagrangian without couplings to soft fields is equivalent to the original QCD Lagrangian \cite{Beneke:2002ph}, we use standard QCD Feynman rules for interactions and quark propagators, respectively.

\section{Calculation of the bare jet function}
\label{sec:barecalc}
The bare jet function, as the superscript suggests, can be expanded in the strong coupling constant as
\begin{equation}
	\label{eq:expansion}
	\begin{aligned}
		J^{(0)}_g(p^2)=1+\frac{Z_{\alpha_s}\alpha_s}{4\pi}J_{1,g}^{(0)}(p^2)+\left(\frac{Z_{\alpha_s}\alpha_s}{4\pi}\right)^2J_{2,g}^{(0)}(p^2)+\mathcal{O}(\alpha_s^3).
	\end{aligned}
\end{equation}
In the equation above, $\alpha_s$ is the renormalized strong coupling constant in the $\overline{\text{MS}}$ scheme and $Z_{\alpha_s}$ is given by
\begin{equation}
	\label{eq:Zas}
	\begin{aligned}
		Z_{\alpha_s}=1-\beta_{0} \frac{\alpha_{s}}{4 \pi \epsilon}+\mathcal{O}\left(\alpha_{s}^{2}\right)\,.
	\end{aligned}
\end{equation}
\subsection{One-loop expression}
\begin{figure}[t]
	\begin{center}
		\includegraphics[width=1\textwidth]{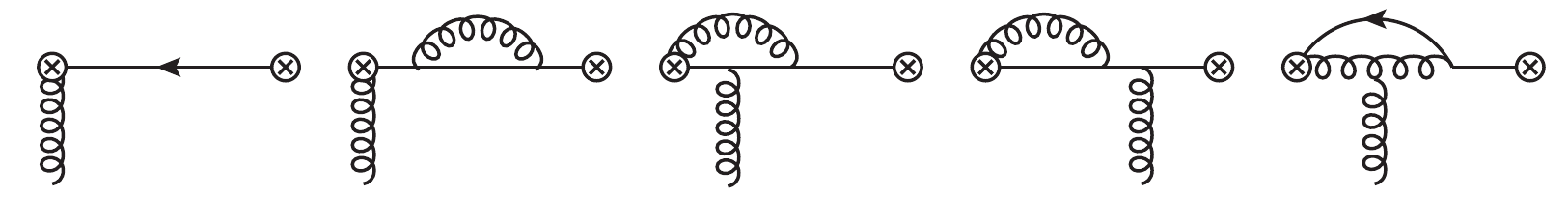}
		\caption{Feynman diagrams contributing to the jet function at one-loop order in light-cone gauge. Note that the third diagram evaluates to zero since it it scaleless.\label{fig:diagrams1loop}}
	\end{center}
\end{figure}

The diagrams contributing at one-loop order are shown in figure \ref{fig:diagrams1loop}. Evaluating these diagrams in $D=4-2\epsilon$ dimensions, we find for the bare gluon jet function up to order $\mathcal{O}(\alpha_s)$ 
\begin{equation}
	\label{eq:bare1loop}
	J^{(0)}_g(p^2)=1+(C_F-C_A)\frac{\alpha_{s,0}}{4\pi}(-p^2-i0)^{-\epsilon}e^{\epsilon\gamma_E}\frac{\Gamma(1+\epsilon)\Gamma^2(-\epsilon)}{\Gamma(2-2\epsilon)}(2-4\epsilon-\epsilon^2)+\mathcal{O}(\alpha_{s,0}^2)\,.
\end{equation}
We find that at one-loop order the gluon jet function is equal to  the photon jet function $J^\gamma(p^2)$ under the replacement of color factors $C_F\to C_F-C_A$.
\subsection{Two-loop expression}
\begin{figure}[t]
	\begin{center}
		\includegraphics[width=1\textwidth]{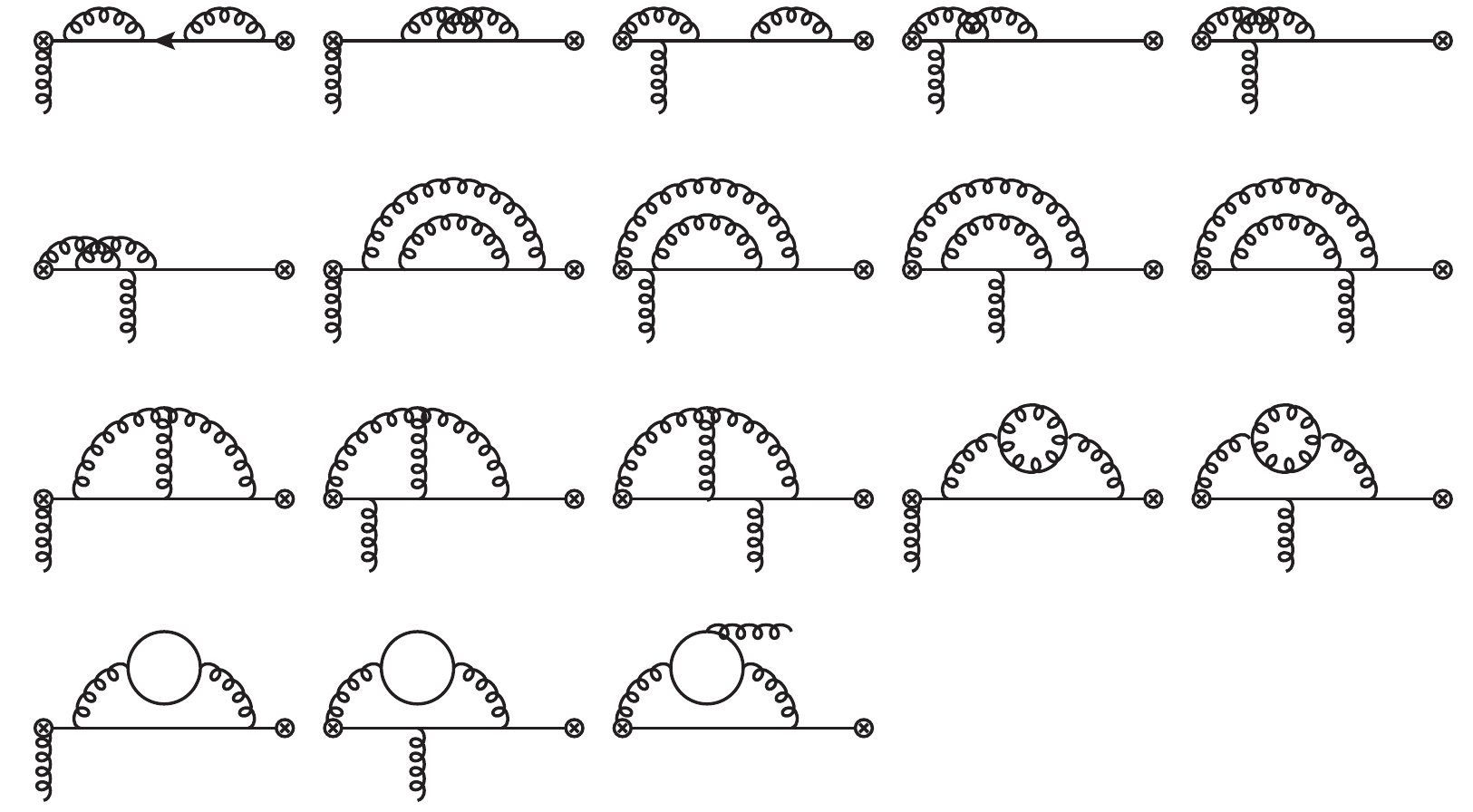}
		\caption{Feynman diagrams contributing to the jet function at two-loop order in light-cone gauge. The diagrams portrayed here also contribute to the photon jet function under the exchange of the external gluon by an external photon. Note that the fermion in the inserted loop can run in both directions (last row).\label{fig:diagrams2loop1}}
	\end{center}
\end{figure}

\begin{figure}[t]
	\begin{center}
		\includegraphics[width=1\textwidth]{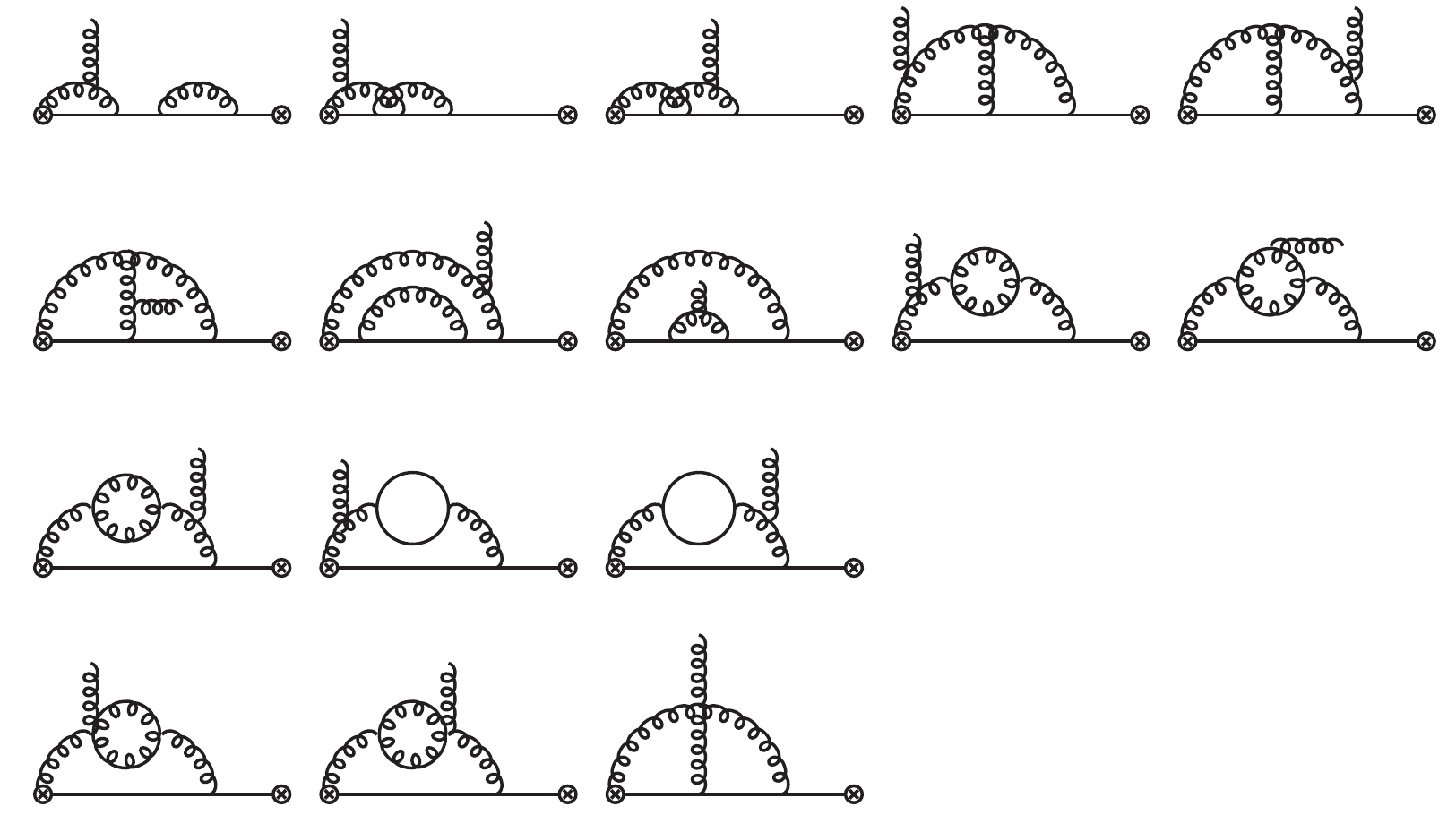}
		\caption{Feynman diagrams contributing to the jet function at two-loop order in light-cone gauge. These diagrams involve three (first rows) and four point interactions (last row) of gluons. As previously, the fermion in the diagrams in the third row can run in both directions.\label{fig:diagrams2loop2}}
	\end{center}
\end{figure}

	At second order in the strong coupling constant $\alpha_s$, the diagrams contributing in light-cone gauge are presented in figures \ref{fig:diagrams2loop1} and \ref{fig:diagrams2loop2}, where we split the diagrams up into those where the external gluon is radiated from a fermion and those where it is radiated from an internal gluon.
Using simplifications of Dirac and Lorentz structures, those diagrams can be brought into a form that consists of linear combinations of scalar Feynman integrals. We now use the integration-by-parts method (IBP) to reduce these Feynman integrals to linear combinations of a limited number of Master Integrals (MI). With the help of kinematic relations we are able to match the found MIs to the ones already encountered in the calculation of the two-loop photon jet function. Calculations were partially performed with \verb+LiteRed+ \cite{Lee:2013mka} and \verb+FIRE6+ \cite{Smirnov:2019qkx}\\
We find for the two-loop contribution to the gluon jet function
\begin{equation}
	\label{eq:total}
	\begin{aligned}
		J_{2,g}^{(0)}(p^2)=e^{-2\epsilon\, L_p}\left[C_F^2K_{FF}+C_FC_AK_{FA}+C_A^2K_{AA}+C_FT_Fn_fK_{Fn_f}+C_AT_Fn_fK_{An_f}\right],
	\end{aligned}
\end{equation}
where $L_p=\ln\frac{-p^2}{\mu^2}$ and the two-loop coefficients are
\begin{equation}
	\begin{aligned}
		K_{FF}=&\frac{2}{\epsilon^4}-\frac{1}{\epsilon^2}\bigg(2+\frac{\pi ^2}{3}\bigg)-\frac{1}{\epsilon}\bigg(4+\frac{\pi ^2}{2}+\frac{46 \zeta_3}{3}\bigg)-\frac{13}{2}-\frac{\pi ^2}{6}-39 \zeta _3+\frac{\pi ^4}{5},\\
		K_{FA}=&-\frac{4}{\epsilon^4}+\frac{11}{6\epsilon^3}+\frac{1}{\epsilon^2}\bigg(\frac{139}{18}+\frac{\pi ^2}{2}\bigg)+\frac{1}{\epsilon}\bigg(\frac{319}{27}-\frac{\pi ^2}{18}+\frac{80 \zeta_3}{3}\bigg)+\frac{1087}{162}-\frac{83 \pi ^2}{54}+\frac{485 \zeta_3}{18}-\frac{49 \pi ^4}{360},\\
		K_{AA}=&\frac{2}{\epsilon^4}-\frac{11}{6\epsilon^3}-\frac{1}{\epsilon^2}\bigg(\frac{103}{18}+\frac{\pi ^2}{6}\bigg)-\frac{1}{\epsilon}\bigg(\frac{485}{54}-\frac{7 \pi ^2}{9}+\frac{34 \zeta_3}{3}\bigg)-\frac{1321}{162}+\frac{56 \pi ^2}{27}+\frac{385 \zeta_3}{18}-\frac{23 \pi ^4}{360},\\
		K_{Fn_f}=&-\frac{2}{3\epsilon^3}-\frac{10}{9\epsilon^2}-\frac{1}{\epsilon}\bigg(\frac{20}{27}-\frac{\pi ^2}{9}\bigg)+\frac{230}{81}+\frac{5 \pi ^2}{27}+\frac{64 \zeta _3}{9},\\
		K_{An_f}=&\frac{2}{3\epsilon^3}+\frac{10}{9\epsilon^2}+\frac{1}{\epsilon}\bigg(\frac{29}{27}-\frac{2 \pi ^2}{9}\bigg)-\frac{131}{81}-\frac{10 \pi ^2}{27}-\frac{106 \zeta_3}{9}.
	\end{aligned}
\end{equation}
Note that in comparison with the photon jet function a simple relation of color factors is not found beyond one-loop order, implying the need for a dedicated renormalization calculation.

\section{Renormalization of the gluon jet function}
\label{sec:reno}
\begin{figure}[t]
	\begin{center}
		\includegraphics[width=1\textwidth]{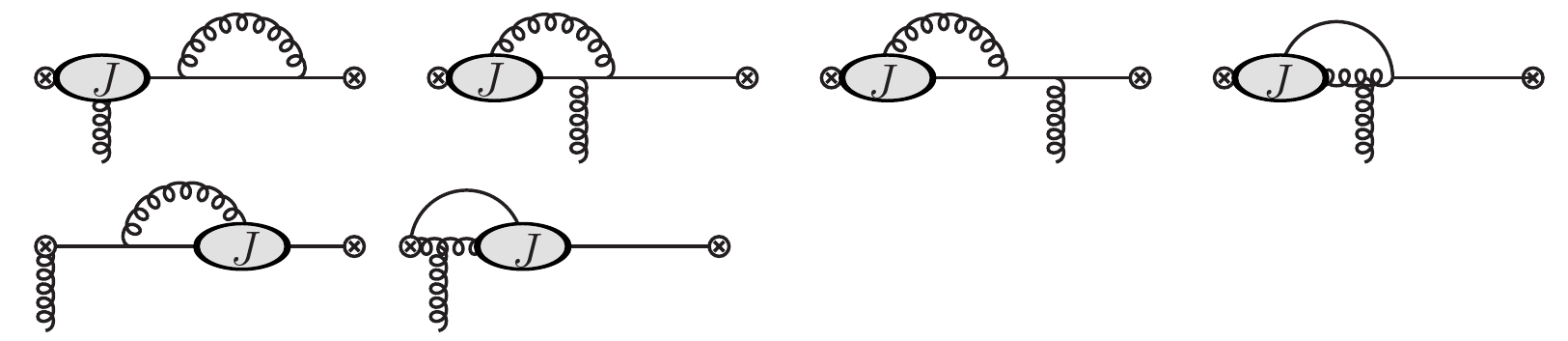}
		\caption{Representative Feynman diagrams contributing to the renormalization of the jet function at one-loop order in light-cone gauge. The blob represents the insertion of the jet function at one-loop order. \label{fig:diagramsreno1}}
	\end{center}
\end{figure}

In the $\overline{\text{MS}}$-scheme, the jet function renormalizes in an integral manner
\begin{equation}
	\label{eq:Renormalizationdef}
	J_g(p^2,\mu)=\int\limits_0^\infty\df x\,Z_{J_g}(p^2,xp^2;\mu)\,J^{(0)}_g(xp^2)\,.
\end{equation}
At one-loop order it would suffice to renormalize the jet function with a local counterterm. However, the correct renormalization factor involves a non-local evolution kernel.\\
The calculation of the renormalization factors is done with two different methods:
\begin{enumerate}
	\item Direct calculation using a technique first presented in  \cite{Bodwin:2021cpx}.
	\item Calculation based on a conjecture of RG consistency in the factorization formula of $gg\to h$ via $b$ quark loop.
\end{enumerate}
As was laid out in \cite{Bodwin:2021cpx}, we can calculate the one-loop renormalization factor of the jet function by consistently inserting the one-loop expression of the jet function into the one-loop diagrams and only keeping the divergent parts. In addition we make use of the kinematic structure of the jet function, i.e. that $J^{(0), n}_g(p^2)\propto(-p^2-i0)^{-n\epsilon}$ at $n$-th order in $\alpha_s$. The diagrams corresponding to this precedure are shown in figure \ref{fig:diagramsreno1}.
We find
\begin{equation}
	\label{eq:ZJNLO}
	Z_{J_g}\left(y p^{2}, x p^{2};\mu\right)=\left[1+\frac{(C_{F}-C_A) \alpha_{s}}{4 \pi}\left(-\frac{2}{\epsilon^{2}}+\frac{2L_p}{\epsilon} \right)\right] \delta(y-x)+\frac{(C_{F}-C_A/2) \alpha_{s}}{2 \pi \epsilon} \Gamma(y, x)\,,
\end{equation}
where $\Gamma(x,y)$ is the Lange-Neubert kernel \cite{Lunghi:2002ju, Bosch:2003fc}. It is a symmetric distribution given by
\begin{equation}
	\label{eq:LNK}
	\Gamma(y,x)=\left[\frac{\theta(y-x)}{y(y-x)}+\frac{\theta(x-y)}{x(x-y)}\right]_+\,.
\end{equation}
The plus-prescription is defined such that when $\Gamma(y,x)$ is to be integrated with a function $f(x)$ one has to replace $f(x)\to f(x)-f(y)$ under the integral.\\
In contrast to the soft quark soft function, whose renormalization was calculated in \cite{Bodwin:2021cpx}, we do not have to deal with non-trivial momentum rooting, further simplifying our calculation. Most recently, the same authors presented their direct calculation of the photon jet function adapting their previously presented method \cite{Bodwin:2021epw}. A more detailed description of our calculation can be found in \cite{Jetfunction}.\\

Our calculation can be further checked by deriving the correct renormalization factor with the help of a conjecture. For this let us first introduce some notations and quantities from the $gg\to h$ factorization theorem. The bare amplitude can be written as 
\begin{equation}
	\label{eq:amplhgg}
	\mathcal{M}(gg\to h)=Z_{gg}^{-1}\big(T_1+T_2+T_3\big)\,,
\end{equation}
where $Z_{gg}^{-1}$ is a global factor in $\overline{\text{MS}}$ scheme accounted for the extra IR divergence \cite{Becher:2009cu,Becher:2009qa} because the initial states carry color charge, and $T_i$ are combinations of hard Wilson coefficient functions and operator matrix elements. $T_3$ is given by
\begin{equation}
	\label{eq:T3}
	T_3=H_3^{(0)}\lim_{\sigma\to-1}\int\limits_0^{M_h}\frac{\df \ell_-}{\ell_-}\int\limits_0^{\sigma M_h}\frac{\df\ell_+}{\ell_+}\,J^{(0)}_g(M_h\ell_-)\,J^{(0)}_g(-M_h\ell_+)\,S^{(0)}(\ell_-\ell_+)\bigg|_\text{leading power}\,,
\end{equation}
where $S^{(0)}$ is the bare soft quark soft function and $H_3^{(0)}$ is the hard matching coefficient, which is the same for the abelian ($h\to\gamma\gamma$) and the non-abelian ($gg\to h$) case. The term $T_3$ is, up to cutoff corrections, RG invariant by itself, which allows us to deduce the renormalization factor of the soft quark soft function given that we conjecture the jet function renormalization factor. Using \eqref{eq:ZJNLO} as the jet function renormalization we find that the deduced soft renormalization factor renormalizes the soft function in a non-trivial way.

\section{Conclusion}
We presented a detailed study of the radiative gluon jet function, which plays a central role in the factorization theorem of the important Higgs production channel $gg\to h$ via a light quark loop. The bare calculation of this object to two-loop order has been demonstrated here for the first time. Additionally, we presented the calculation of the non-trivial renormalization factor at one-loop order, where we used two different and complementary methods.\\
The results of this paper are crucial ingredients in the already mentioned factorization theorem. It serves not only as a way to bootstrap the renormalization of the soft quark soft function, but also helps us to remove endpoint divergences elegantly with the use of refactorization theorems. Furthermore, knowledge of the jet function and it renormalization beyond leading order allows us to predict large logarithmic contributions in the three-loop amplitude of $gg\to h$.\\
This proceeding is the precursor for a detailed article on the subject, see \cite{Jetfunction}.

\section*{Acknowledgements}
The author wishes to thank Matthias Neubert, Xing Wang and Zelong Liu for exciting and fruitful work together on this project. The author would like to thank the organizers of RADCOR 2021 for the well-organized conference.

\paragraph{Funding information}
This research was supported by the Cluster of Excellence \textit{Precision Physics, Fundamental Interactions and Structure of Matter} (PRISMA$^+$ - EXC 2118/1) within the German Excellence Strategy (project ID 39083149).



\bibliography{references.bib}

\nolinenumbers

\end{document}